\documentclass[aps,prd,twocolumn,nofootinbib,superscriptaddress,floatfix,notitlepage]{revtex4-1}

\usepackage{graphicx}
\usepackage{amsmath,amsfonts,amssymb}
\usepackage{color}
\usepackage[breaklinks,colorlinks,urlcolor=blue,citecolor=blue,linkcolor=magenta]{hyperref}
\usepackage{verbatim}
\usepackage{multirow}

\newcommand{\be}{\begin{equation}} 
\newcommand{\ee}{\end{equation}}
\newcommand{\bea}{\begin{equation}\begin{aligned}} 
\newcommand{\eea}{\end{aligned}\end{equation}}
\newcommand{\td}{{\rm d}}

\begin{document}

\title{Gravitational waves from bubble collisions and fluid motion \\in strongly supercooled phase transitions}

\author{Marek Lewicki}
\email{marek.lewicki@fuw.edu.pl}
\affiliation{Faculty of Physics, University of Warsaw ul.\ Pasteura 5, 02-093 Warsaw, Poland}
\author{Ville Vaskonen}
\email{vvaskonen@ifae.es}
\affiliation{Institut de Fisica d'Altes Energies, Campus UAB, 08193 Bellaterra (Barcelona), Spain}

\begin{abstract}
We estimate the gravitational wave spectra generated in strongly supercooled phase transitions by bubble collisions and fluid motion. We derive analytically in the thin-wall approximation the efficiency factor that determines the share of the energy released in the transition between the scalar field and the fluid. We perform numerical simulations including the efficiency factor as a function of bubble radius separately for all points on the bubble surfaces to take into account their different collision times. We find that the efficiency factor does not significantly change the gravitational wave spectra and show that the result can be approximated by multiplying the spectrum obtained without the efficiency factor by its value at the radius $R_{\rm eff} \simeq 5/\beta$, where $\beta$ is the approximate inverse duration of the transition. We also provide updated fits for the gravitational wave spectra produced in strongly supercooled transitions from both bubble collisions and fluid motion depending on the behaviour of the sources after the collision.
\end{abstract}

\maketitle

\section{Introduction}

The first observations of gravitational waves (GWs) by LIGO/Virgo signified the beginning of a new era in astrophysics and cosmology. While up to now all observed events were produced by compact object binaries~\cite{LIGOScientific:2018mvr,LIGOScientific:2020ibl,LIGOScientific:2021usb,LIGOScientific:2021djp}, this new messenger brings hope also for detection of primordial signals in the form of stochastic GW backgrounds. Given the tremendous advancements in sensitivity that are expected throughout a broad frequency spectrum with the upcoming experiments~\cite{Punturo:2010zz,Hild:2010id,Janssen:2014dka,Graham:2016plp,LISA:2017pwj,Graham:2017pmn,Badurina:2019hst,AEDGE:2019nxb,Bertoldi:2021rqk,Alonso:2022oot,Badurina:2021rgt}, the prospects for probing the early Universe processes are great even though the compact object binaries that will contribute to the stochastic GW background make the detection of its primordial components more difficult~\cite{Lewicki:2021kmu}. Interestingly, the recent pulsar timing observations~\cite{NANOGrav:2020bcs,Goncharov:2021oub,Chen:2021rqp,Antoniadis:2022pcn} feature a common spectrum-process which could be an early indication of the upcoming first detection of a stochastic GW background, potentially of primordial origin~\cite{Ellis:2020ena,Blasi:2020mfx,Vaskonen:2020lbd,DeLuca:2020agl,Nakai:2020oit,Ratzinger:2020koh,Kohri:2020qqd,Vagnozzi:2020gtf,Neronov:2020qrl,Blanco-Pillado:2021ygr,Wang:2022wwj,RoperPol:2022iel,Ferreira:2022zzo}.     

Many high-energy processes, including phase transitions~\cite{Caprini:2015zlo,Caprini:2019egz}, cosmic strings~\cite{Auclair:2019wcv} and inflation~\cite{Bartolo:2016ami}, occurring in the early Universe may generate a detectable stochastic GW background. In this paper we focus on cosmological first-order phase transitions featured in various particle physics models. They are intensive processes where bubbles of the new phase nucleate, expand and eventually convert the whole Universe in the true vacuum phase~\cite{Coleman:1977py}. Interactions between the expanding bubble walls and the surrounding fluid cause motion and inhomogeneities in the fluid, and both the collisions of the bubble walls and the motion of fluid inhomogeneities source GWs~\cite{Kosowsky:1992vn,Kamionkowski:1993fg}. The resulting GW spectra from these components have been extensively studied with numerical and semi-analytical methods (see e.g.~\cite{Hindmarsh:2019phv,Cutting:2019zws,Lewicki:2020jiv,Lewicki:2020azd,Jinno:2020eqg,Dahl:2021wyk,Cutting:2022zgd} for recent progress). These studies indicate that different sources active during the transition can produce different GW spectra.

In order to determine the GW spectrum generated in a phase transition in a given particle physics model, we need to estimate how much each of the GW sources contributes to the final GW spectrum. The vacuum energy released in the transition is split between the gradient energy of the scalar field bubble wall and motion in the fluid. How the total released energy is split depends on the strength of the interactions between the wall and the particles in the fluid, and on the strength of the transition. 

In strongly supercooled phase transitions it is possible that the interactions of the bubble wall with fluid do not stop the wall from accelerating before it collides with other bubbles. In this case most of the released energy is in the bubble walls and the bubble collisions give the dominant contribution to the GW spectrum. This can happen in particular in quasi-conformal models~\cite{Jinno:2016knw,Iso:2017uuu,Marzola:2017jzl,Prokopec:2018tnq,Marzo:2018nov,Baratella:2018pxi,VonHarling:2019rgb,Aoki:2019mlt,DelleRose:2019pgi,Wang:2020jrd,Ellis:2020nnr,Baldes:2020kam,Baldes:2021aph,Lewicki:2021xku}. If the interactions instead are sufficiently strong, the bubble wall reaches a terminal velocity before the collisions and majority of the released energy goes into fluid motion. This is the typical case in extensions of the Standard Model featuring polynomial scalar potentials~\cite{Grojean:2006bp,Dorsch:2014qpa,Huang:2016cjm,Artymowski:2016tme,Vaskonen:2016yiu,Dorsch:2016nrg,Ellis:2018mja,Beniwal:2018hyi,Fairbairn:2019xog,Ellis:2019oqb,Lewicki:2021pgr}.

In this paper we derive analytically an efficiency factor that determines how large is the contribution from each of the GW sources. We perform numerical simulations of the phase transition, describing both of the GW sources, bubble walls and fluid motion, in the thin-wall limit, to show how the efficiency factor affects the final GW spectrum. Moreover, we derive analytically the probability density function for the radius at which a given point on the surface of a bubble collides with another bubble and verify the results against our numerical simulations. Finally we also provide updated fits to the spectral shapes of the GW signals that  can be produced by all sources active in very strong phase transitions.

\section{Energy budget}

We estimate how the energy released in the bubble expansion is shared between the scalar field gradients and the fluid motion in strongly supercooled phase transitions by studying the bubble expansion under the influence of pressure terms caused by the interactions of the wall with the ambient fluid. We perform the computation consistently in the thin-wall limit, which gives a good description of the system if the bubble reaches ultra-relativistic velocities. The following analysis improves earlier approximations used in the literature~\cite{Ellis:2019oqb,Ellis:2020nnr,Cai:2020djd}.

In the thin-wall limit the energy carried by the bubble walls can be modeled using a simple analytical prescription. This assumes that the bubble walls are spherical shells with a certain surface energy density and the interactions of the walls with the ambient fluid are local. In this limit, neglecting the expansion of the Universe,\footnote{The assumption of neglecting the expansion of the Universe is in the end related to the nucleation rate of bubbles, and it is valid if the bubble radius at collision moment is much smaller than the Hubble radius, $\langle R_c \rangle \ll 1/H$, which translates to the requirement that $\beta/H \gg 1$ (see Sec.~\ref{sec:nucleation}.) } the evolution of the bubble radius $R$ can be described by 
the equation of motion~\cite{Lewicki:2022nba}
\be \label{eq:eomR}
\ddot R + \frac{2}{R} (1-\dot R^2) = \frac{\Delta P(\dot R)}{\sigma} (1-\dot R^2)^{3/2}
\ee
that arises from energy conservation of the coupled system of fluid and the scalar field bubble. Here $\Delta P(\dot R)$ denotes the pressure difference across the bubble wall and $\sigma$ is the surface tension of the wall. The latter is defined through the scalar potential $V$ as~\cite{Coleman:1977py}
\be
\sigma \equiv \int_0^{\varphi_c} \td \varphi \sqrt{2V(\varphi)} \,,
\ee
where $\varphi_c>0$ denotes the field value at which the potential energy is the the same as in the false vacuum that lies at the origin, $V(\varphi_c) = V(0)$. 

In terms of the Lorentz factor of the bubble wall, $\gamma = 1/\sqrt{1-\dot R^2}$, the equation of motion~\eqref{eq:eomR} is given by
\be \label{eq:eom}
\frac{\td\gamma}{\td R} + \frac{2\gamma}{R} = \frac{\Delta P(\gamma)}{\sigma} \,.
\ee
The bubble nucleates at rest, $\gamma=1$, $\td \gamma/\td R = 0$, with an initial radius $R_0$. By Eq.~\eqref{eq:eom} we can relate the wall tension to the initial radius as $R_0 = 2\sigma/\Delta P_0$, where $\Delta P_0 \equiv \Delta P(\gamma=1)$. For a constant pressure difference, $\Delta P = \Delta P_0$, the solution of the equation of motion is
\be \label{eq:gammaequ}
\gamma = \frac{2 R}{3 R_0} + \frac{R_0^2}{3 R^2} \,.
\ee

\begin{figure*}
\centering
\includegraphics[width=\textwidth]{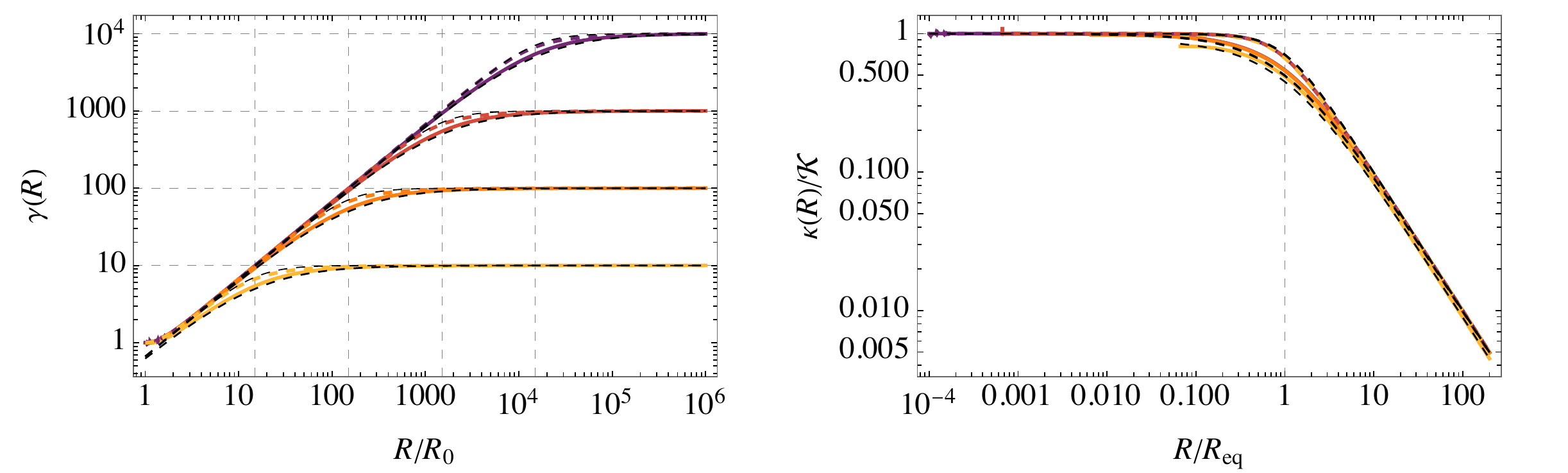}
g\caption{The Lorentz factor of the bubble wall (left panel) and the efficiency factor $\kappa$ (right panel) as a function of the bubble radius $R$ for $\gamma_{\rm eq} = 10,100,10^3,10^4$ from light to dark. The solid black curves show the case $\Delta P_{1\to N}\propto \gamma$ and the orange dashed curves the case $\Delta P_{1\to N}\propto \gamma^2$. In the left panel the vertical dashed lines indicate the values of $R_{\rm eq}/R_0$. The colored curves show the full solution of the equation of motion~\eqref{eq:eom}, while the black dashed curves correspond to the approximations~\eqref{eq:gammaapr} and \eqref{eq:kappaapr}.}
\label{fig:gamma}
\end{figure*}

The total pressure difference across the bubble wall, accounting for $1\to 1$ scatterings and $1\to N$ splittings at the bubble wall, is given by
\be
\Delta P(\gamma) = \Delta V - \Delta P_{1\to 1} - \Delta P_{1\to N}(\gamma) \,,
\ee
where $\Delta V$ denotes the potential energy difference between the minima. The pressure arising from $1\to 1$ scatterings quickly reaches a constant value in the relativistic limit~\cite{Bodeker:2009qy,Lewicki:2022nba}. Subsequently, the $\gamma$ dependence of the total pressure difference arises only from $\Delta P_{1\to N}$, for which we consider two forms. The first one, suggested in~\cite{Bodeker:2017cim,Azatov:2020ufh,Gouttenoire:2021kjv} is linear in the Lorentz factor $\Delta P_{1\to N} = \tilde \Delta P_{1\to N} \gamma$, and the second one, suggested in~\cite{Hoche:2020ysm,BarrosoMancha:2020fay}, is quadratic in the Lorentz factor, $\Delta P_{1\to N} = \tilde \Delta P_{1\to N} \gamma^2$ In both cases $\tilde \Delta P_{1\to N}$ is a constant. 

By plugging $\Delta P(\gamma)$ into~\eqref{eq:eom}, we can easily solve $\gamma$ as a function of $R$. When $\Delta P_{1\to N} \ll \Delta V - \Delta P_{1\to 1}$ the solution can be approximated by Eq.~\eqref{eq:gammaequ}. Assuming in addition that $R\gg R_0$, the Lorentz factor grows linearly with the radius, $\gamma \approx 2R/(3R_0)$. Eventually, as the bubble wall accelerates, $\gamma$ becomes large enough for the $1\to N$ splittings to be important, $\Delta P_{1\to N} \sim \Delta V - \Delta P_{1\to 1}$, after which it asymptotically reaches the value
\be \label{eq:gammaeq}
\gamma_{\rm eq} \equiv \left[\frac{\Delta V - \Delta P_{1\to 1}}{\tilde \Delta P_{1\to N}} \right]^{\frac1c} \,,
\ee
where $c=1,2$ depending on the scaling of the $1\to N$ pressure, $\Delta P_{1\to N} \propto \gamma^c$. The change from the linear growth to asymptotically constant behaviour occurs when the radius reaches 
\be \label{eq:Req}
R_{\rm eq} \approx \frac32 R_0 \gamma_{\rm eq} \,.
\ee
The solution $\gamma(R)$, in the limit $\gamma_{\rm eq}\gg1$, can be approximated by a simple broken power-law
\be \label{eq:gammaapr}
\gamma(R) = \gamma_{\rm eq} \left[ 1+ \left(\frac{R_{\rm eq}}{R}\right)^{\!\!c}\,\right]^{-1/c} \,.
\ee
In Fig.~\ref{fig:gamma} we show the full solution $\gamma(R)$ for different values of $\gamma_{\rm eq}$ for both $\Delta P_{1\to N}\propto \gamma$ (solid) and $\Delta P_{1\to N}\propto \gamma^2$ (dashed) together with the above approximation shown by the dotted black curves. In both cases the transition from linear growth, $\gamma\propto R$ to the constant value $\gamma \approx \gamma_{\rm eq}$ is quite fast, and the difference between the two cases is small. The main effect of the scaling of $\Delta P_{1\to N}$ is that it changes $\gamma_{\rm eq}$ and $R_{\rm eq}$.

We define the efficiency factor $\kappa$ as the fraction of the total released energy within a unit solid angle that goes into the bubble wall energy, 
\be
\kappa(R) = \frac{3(\gamma R^2 -R_0^2)\sigma}{(R^3 - R_0^3) \Delta V} \,.
\ee
The rest of the released energy, $1-\kappa(R)$, goes into fluid motion. This is a good approximation for strongly supercooled transitions. In weaker transitions one also needs to keep in mind that some of the energy going into the fluid will be lost on its heating which will reduce the overall GW signal from the fluid motion~\cite{Espinosa:2010hh,Ellis:2019oqb}. 

Using the approximation~\eqref{eq:gammaapr}, we can express the efficiency factor as
\be \label{eq:kappaapr}
\kappa(R) \approx \mathcal{K} \, \frac{R_{\rm eq}}{R} \frac{\gamma(R)}{\gamma_{\rm eq}} ,
\ee
where 
\be \label{eq:K}
\mathcal{K} \equiv \left[1-\frac{\alpha_\infty}{\alpha}\right] \left[ 1-\frac{1}{\gamma_{\rm eq}^c} \right] 
\ee
is a constant, $\mathcal{K} < 1$. The parameters $\alpha$ and $\alpha_\infty$ are defined by scaling with the radiation energy density $\rho_R$ as $\alpha = \Delta V/\rho_R$ and $\alpha_\infty = \Delta P_{1\to 1}/\rho_R$ (see~\cite{Ellis:2019oqb} for more details). Typically for strongly supercooled transitions $\mathcal{K} \approx 1$. As shown in the right panel of Fig.~\ref{fig:gamma}, the efficiency factor remains constant at $R\ll R_{\rm eq}$ and decreases as $\kappa\propto 1/R$ at $R\gg R_{\rm eq}$. In the same way as for $\gamma(R)$, the difference between the cases $\Delta P_{1\to N}\propto \gamma$ and $\Delta P_{1\to N}\propto \gamma^2$ is small.

\section{Bubble nucleation and collisions}
\label{sec:nucleation}

Soon after the bubble has nucleated, we can neglect its initial radius, and, if the friction terms are sufficiently small ($\gamma_{\rm eq}\gg 1$), we can approximate that the bubble radius grows as $R = t-t_n$, where $t_n$ denotes the nucleation time of the bubble. Moreover, assuming that the bubbles are much smaller than the Hubble horizon, we can neglect the expansion of the Universe. The expected number of bubbles reaching a given point is then given by
\be
N(t) =  \frac{4\pi}{3}  \int_{-\infty}^t \!\! \td t' (t-t')^3 \Gamma(t') \,,
\ee
where $\Gamma(t)$ denotes the bubble nucleation rate per unit time and volume, and the probability that the given point still is in the false vacuum at time $t$ is
\be
P(t) = e^{-N(t)} \,.
\ee

Let us consider a bubble nucleation rate $\Gamma(t) = C e^{A(t)}$. Around the time $t_*$ when the transition proceeds, we can expand $A(t)$ to get $\Gamma(t) = C e^{A(t_*) + \beta (t-t_*)} = \Gamma_0 e^{\beta t}$, where $\beta \equiv \td \ln\Gamma/\td t |_{t=t_*}$ and $\Gamma_0 \equiv C e^{A(t_*) - \beta t_*}$. As the transition is not an instantaneous process, the choice of $t_*$ includes some ambiguity. It is convenient to choose $t_*$ by requiring that $P(t_*) = 1/e$, which gives $\Gamma_0 = \beta^4/(8\pi)$, and $N(t) = e^{\beta t}$. 

Next, let us consider a point on the surface of a bubble that nucleated at time $t_n$. If the point is still in the false vacuum when the radius of the bubble is $R$, then it has stayed in the false vacuum for the whole time $0 \leq t-t_n < R$. The probability for this is $P(t_n + R)$. So, the probability that a bubble nucleated within time $t_n < t < t_n + \td t_n$ in a volume $V$, and that a point on its surface is still in the false vacuum at radius $R$, is given by $\td t_n \,V \,\Gamma(t_n) P(t_n + R)$. By integrating this over the nucleation time $t_n$ we get the probability density function for the radius at which a bubble surface element collides with the surface of another bubble,
\be \label{eq:pRc}
p(R_c) \propto \int \td t_n \, \Gamma(t_n) P(t_n + R_c) \,,
\ee
which we normalize to unity, $\int \td R_c \,p(R_c) = 1$. For the exponential bubble nucleation rate, $\Gamma(t) \propto e^{\beta t}$, this gives (independently of the prefactor $\Gamma_0$)\footnote{This agrees with the distribution on the bubble lifetime derived in Ref.~\cite{Hindmarsh:2019phv}. Our result can be generalized to wall velocities $v_w < 1$ simply by dividing $R_c$ by $v_w$ in the exponent.}
\be \label{eq:pRcexp}
p(R_c) = \beta e^{-\beta R_c} \,.
\ee

\begin{figure}
\centering
\includegraphics[width=\columnwidth]{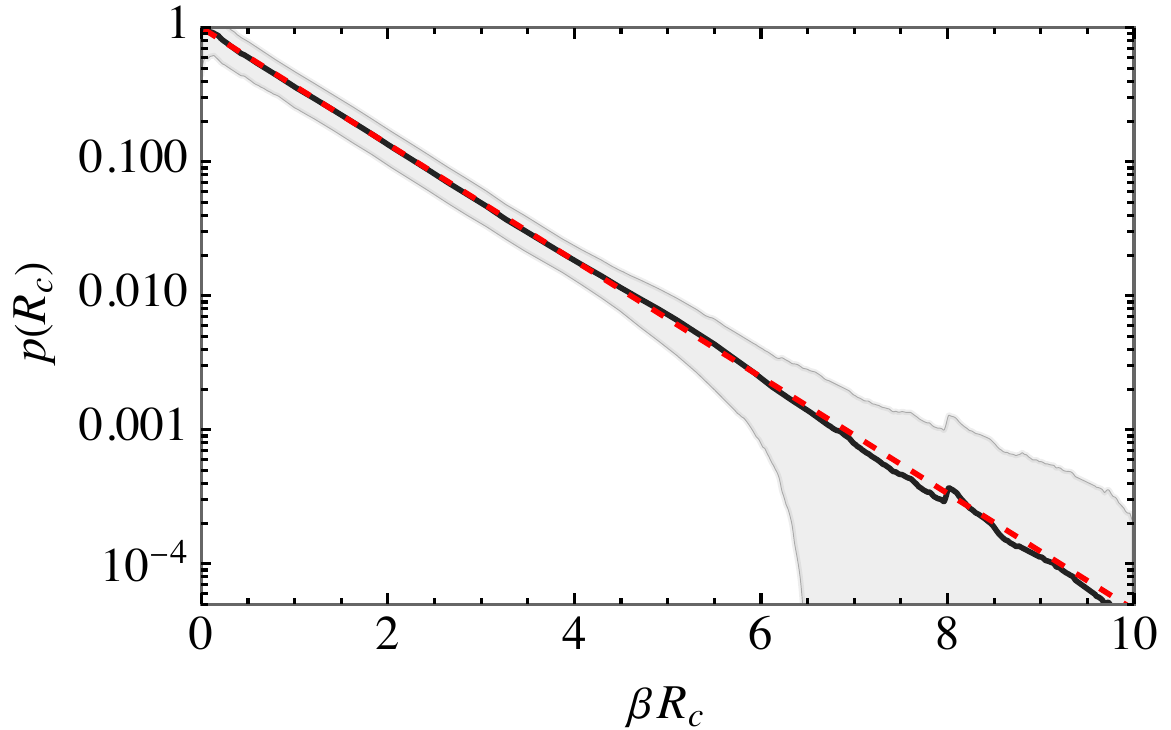}
\caption{Probability distribution function of bubble radius at collision. The red dashed curve shows the analytical result~\eqref{eq:pRcexp}. The black curve and the gray band show the mean and the variance of the result from our numerical simulations of bubble nucleation.}
\label{fig:pRc}
\end{figure}

The above result provides a good cross-check for the numerical simulations that we will use for the GW computation. In Fig.~\ref{fig:pRc} the solid curve and the gray band indicate the mean and variance of the $R_c$ distribution obtained from 90 simulations with simulation volume $(16/\beta)^3$ and each of the simulation including at least 70 bubbles. In these simulations we nucleate thin-wall bubbles according to the exponential rate inside a cubic box with periodic boundary conditions, evolve them according to $R = t-t_n$, discretise the bubble surfaces, and find the radius at which each of the points on the bubble surface collide with a wall of another bubble using the cosine rule. We label the bubbles with index $j$ and denote the position vectors of the bubble centers by $\vec{x}_j$. Consider a point defined by the angles $\theta$ and $\phi$ on the surface of the bubble $j=j'$. The radius at which that point collides with a surface of another bubble is given by
\be \label{eq:Rc}
R_{c} = \min_{j\neq j'}\left[\frac{d_{j}^2 - \Delta t_{j}^2}{2(d_{j} \cos\theta_{j} - \Delta t_{j})} \right] \,, 
\ee
where the minimum is taken over all bubbles, $d_{j}^2 \equiv |\vec{x}_j - \vec{x}_{j'}|^2$ is the distance between the bubble nucleation centers, $\Delta t_{j} \equiv t_{n,j} - t_{n,j'}$ is the time between their nucleation, and $\theta_{j}$ is the angle between the vector $\vec{x}_j - \vec{x}_{j'}$ and the vector corresponding to the angles $\theta$ and $\phi$. As shown in Fig.~\ref{fig:pRc}, the simulation result agrees well with the analytical result~\eqref{eq:pRcexp}.

A widely used approximation for the bubble radius upon collision comes from the bubble number density $n_{\rm bubbles} = \int \td t_n \Gamma(t_n) P(t_n)$ which leads to $ R_* = n_{\rm bubbles}^{-1/3} = (8\pi)^{1/3}/\beta$. From $p(R_c)$ we can calculate moments of the bubble radius when a bubble surface element collides with the surface of another bubble, $\langle R_c^n\rangle = \int \td R_c\, R_c^n p(R_c)$. For the exponential bubble nucleation rate this gives
\be \label{eq:Rcn}
\langle R_c^n \rangle = n! \beta^{-n} \,.
\ee
Given that the released energy scales with the radius to the third power, this leads to a different estimate of the average bubble radius $\langle R_c^3 \rangle^{1/3} = 6^{1/3}/\beta$ more appropriate for computation of the GW spectrum.

\section{Gravitational waves}

The energy released in the bubble expansion is divided between the bubble wall and the fluid shell that follows right behind the wall. Both the bubble walls and the fluid shells source GWs. We model these sources in the thin-wall limit and calculate the GW spectrum accounting for the efficiency factor $\kappa(R)$ for the bubble collisions and $1-\kappa(R)$ for the fluid motion. The modeling of the fluid motion in the thin-wall limit is based on the assumption that the released energy going to fluid motion is strongly localized in a thin shell. Before collision this fluid shell is right behind the bubble wall and after the collision it propagates to the same direction as before the collision,however , depending on how strong the interaction are between the fluid and the scalar field, it's velocity can slow down to the speed of sound. 

We calculate the GW spectrum as e.g. in Refs.~\cite{Lewicki:2020jiv,Lewicki:2020azd} assuming that, as in the previous section, the bubble nucleation follows exponential rate per unit volume, $\Gamma \propto e^{\beta t}$. Each of the contributions ($l = $bubbles, fluid) to the resulting energy spectrum of GWs can be expressed as
\be \label{eq:Omega}
\Omega_{{\rm GW},l}(f) = \left[\frac{H}{\beta}\right]^2\left[\frac{\alpha}{1+\alpha}\right]^2 \,S_l(f) \,,
\ee
where 
\be \label{eq:S}
S_l(f) \!=\! \left(\frac{2\pi f}{\beta}\right)^3 \frac{3\beta^5}{2 V_s} \int \!\frac{\td\Omega_k}{4\pi} \left[ |C_{l,+}(f)|^2 + |C_{l,\times}(f)|^2 \right]
\ee
encodes the spectral shape of the signal. The integral is over the wavevector $\vec{k}$ directions, and the integrand is $\propto V_s/\beta^5$ if the volume $V_s$ over which we average the GW energy spectrum is sufficiently big. 

Using the thin-wall limit, the functions $C_{l,+}$ and $C_{l,\times}$ in the direction $\hat{k} = (0,0,1)$, can be expressed as
\bea \label{eq:Cpc}
C_{l,+,\times}(f) \approx \frac{1}{6\pi} \sum_j &\int_{t_{n,j}} \!\td t\, \td \Omega\, \sin^2\theta\, g_{+,\times}(\phi) \\ &\times R_j^3 f_l(R_j) \,e^{i 2\pi f (t - z_j - R_j\cos\theta)} \,.
\eea
The sum runs over all the bubbles nucleated in the volume $V_s$, $t_{n,j}$ is the time of nucleation of the bubble $j$, $z_j$ is the $z$ coordinate of its center, and $R_j = v_l (t-t_{n,j})$, where $v_l$ is the bubble wall/fluid shell velocity, denotes the radius of the bubble/fluid shell $j$ at time $t$. For the bubble walls we use $v_{\rm bubbles}=1$ both before and after the collision, whereas for the fluid shells we use $v_{\rm fluid}=1$ before the collision and after the collision we consider two cases: $v_{\rm fluid} = 1$ and $v_{\rm fluid} = c_s = 1/\sqrt{3}$. The former is appropriate for very strong transitions~\cite{Jinno:2019jhi}, whereas the latter is realized for weaker transitions~\cite{Jinno:2020eqg}. The functions $g_{+,\times}$ read $g_+(\phi) = \cos(2\phi)$ and $g_\times(\phi) = \sin(2\phi)$. 

The function $f_l(R)$ encodes the scaling of the GW source~\cite{Lewicki:2020azd}. For the bubble collisions contribution, we follow the results of lattice simulations~\cite{Lewicki:2020jiv,Lewicki:2020azd}, which showed that the maximum of the stress-energy tensor scales as $T_{rr} \propto R^{-\xi}$ after the collision. The power $\xi$ in general depends on the underlying particle physics model. In particular, it was shown in~\cite{Lewicki:2020jiv} that breaking of a global symmetry corresponds to $\xi=2$ while in~\cite{Lewicki:2020azd} it was shown that in models where the phase transition breaks a gauge symmetry correspond to $\xi=3$. Accounting also for the efficiency factor $\kappa$, the $f_l$ function for bubble collisions is given by
\be \label{eq:Edecay} 
f_{\rm bubbles}(R) = 
\begin{cases}
\kappa(R) \,, & R \leq R_c \,, \\
\kappa(R_c) \left[\frac{R_c}{R}\right]^{\xi+1} \,, & R > R_c \,,
\end{cases}
\ee
where $R_c$ denotes the bubble radius at the moment of collision, $t=t_c$. In contrast with Refs.~\cite{Lewicki:2020jiv,Lewicki:2020azd}, where $R_c$ was determined numerically by the bisection method, we find $R_c$ using Eq.~\eqref{eq:Rc}.

\begin{figure}
\centering
\includegraphics[width=\columnwidth]{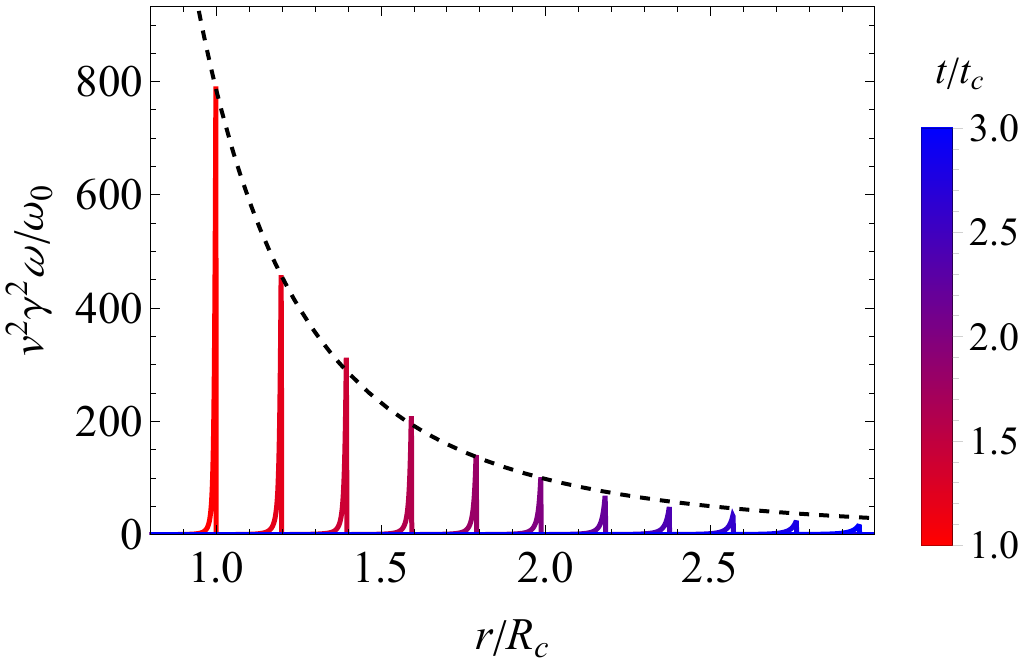}
\caption{Time evolution of the perfect fluid stress energy tensor (solid lines) together with the $r^{-3}$ scaling for comparison (dashed line). The red profile is the initial condition just after the bubble collision $t_c$ and darker colours show the profile at later times. The parameters for this example profile are $\alpha=20$ and $\gamma_w=50$ corresponding to a very strong transition such that the velocity of the profile remains nearly constant and only very slowly changes to the speed of sound.}
\label{fig:Efluid}
\end{figure}

Also for the fluid motion we assume that the maximum of the stress-energy tensor scales as $R^{-\xi}$ after the collision. The function $f_l$ for fluid motion then reads
\be \label{eq:Edecay2}
f_{\rm fluid}(R) = 
\begin{cases}
1-\kappa(R) \,, & R \leq R_c \,, \\
\left[1-\kappa(R_c)\right] \left[\frac{R_c}{R}\right]^{\xi+1} \,, & R > R_c \,.
\end{cases}
\ee 
In the perfect fluid description, that assumes the fluid to remain in local equilibrium at all times, the transverse-traceless part of the stress energy tensor of the fluid reads $T_{ij} = \gamma^2 v_i v_j w$, where $\vec{v}$ is the fluid velocity and $w$ is its enthalpy density. By the interactions of the fluid with the wall, an overdense fluid shell with radial velocity $v_r > 0$ builds up around the bubble wall. If the wall reaches a terminal velocity, the fluid shell settles into a self-similar profile~\cite{Espinosa:2010hh}. We expect that this shell continues to propagate after the bubble wall collides with the wall of another bubble without changing its shape significantly in the collision. This behaviour was first observed for weaker transitions, $\alpha < \mathcal{O}(0.1)$, in lattice simulations~\cite{Hindmarsh:2017gnf,Hindmarsh:2015qta}. For our case of very strong transitions, $\alpha \geq \mathcal{O}(10)$, we used a simplified simulation involving only the fluid and assuming extra symmetry of the system to retain only one spatial dimension as in~\cite{Jinno:2020eqg}.\footnote{We use the integration scheme devised in~\cite{KURGANOV2000241} in order to reduce the numerical diffusion. The parameters used in the simulations are $\Delta x=10^{-4}R_c$ for the step in radius and a tenth of that in time. We also verified reducing the step sizes by an order of magnitude does not modify the results.} We begun with simulating collisions of two planar shells and verified that they are not significantly modified and instead simply propagate onward. We next simulated the evolution of spherically symmetric fluid shells after the collision. Fig.~\ref{fig:Efluid} shows an illustrative example of our results. We find that the maximum of the $rr$ component of the stress energy tensor scales as $T_{rr} \propto R^{-3}$. This matches to the same scaling found in~\cite{Jinno:2020eqg} for weak transitions and motivates us to consider $\xi=3$ for the scaling of the fluid related GW source. For comparison, we consider also $\xi = 2$ which corresponds to the bulk flow model~\cite{Konstandin:2017sat}.

In principle, after the GW generation through relativistic fluid shells finishes, one would expect to enter the period of sound waves~\cite{Hindmarsh:2015qta,Hindmarsh:2016lnk,Hindmarsh:2017gnf,Hindmarsh:2019phv} and perhaps also turbulence~\cite{RoperPol:2019wvy,Kahniashvili:2020jgm,RoperPol:2021xnd,Auclair:2022jod}. However, in the very strong transitions, that are our primary interest, we expect that the fluid will remain in the relativistic shells for a long time after the transition (at least until the shell radius has grown by $\mathcal{O}(1)$ factor). The energy carried by the inhomogeneities has then significantly diluted once the sound wave and turbulence periods begin and, therefore, we expect the main contribution on the GW spectrum in very strong transitions to arise from the relativistic fluid shells or the scalar field bubbles themselves. Thus, in the present analysis we neglect the periods of sound waves and turbulence.

\begin{table*}
\centering
\begin{tabular}{@{\extracolsep{8pt}} lccccccc @{}}
& \multicolumn{3}{c}{Bubbles} & \multicolumn{4}{c}{Fluid} \\
\cline{2-4} \cline{5-8} \\[-9pt]
& \multirow{2}{*}{envelope} &  \multirow{2}{*}{$T_{rr}\propto R^{-2}$} &  \multirow{2}{*}{$T_{rr}\propto R^{-3}$} &  \multicolumn{2}{c}{$T_{rr}\propto R^{-2}$} & \multicolumn{2}{c}{$T_{rr}\propto R^{-3}$} \\
\cline{5-6} \cline{7-8} 
& & & &  $v_{\rm fluid}=1$  &  $v_{\rm fluid}=c_s$ &  $v_{\rm fluid}=1$ &  $v_{\rm fluid}=c_s$ \\
\hline
$100\,A$ & $3.78 \pm 0.04$ & $5.93\pm 0.05$ & $5.13\pm 0.05$ & $5.94\pm 0.02$ & $3.36\pm 0.01$ & $5.14\pm 0.04$ & $3.64\pm 0.02$ \\
$a$ & $3.08 \pm 0.04$ & $1.03\pm 0.04$ & $2.41\pm 0.10$ & $1.03\pm 0.05$ & $1.00\pm 0.05$ & $2.36\pm 0.09$ & $2.02\pm 0.08$ \\
$b$ & $0.98 \pm 0.05$ & $1.84\pm 0.17$ & $2.42\pm 0.11$ & $1.87\pm 0.18$ & $1.39\pm 0.15$ & $2.36\pm 0.09$ & $1.38\pm 0.06$ \\
$c$ & $1.91 \pm 0.29$ & $1.45\pm 0.34$ & $4.08\pm 0.77$ & $1.39\pm 0.38$ & $0.71\pm 0.26$ & $3.69\pm 0.48$ & $1.48\pm 0.32$ \\
$2\pi f_p/\beta$ & $1.33 \pm 0.19$ & $0.64\pm 0.09$ & $0.77\pm 0.12$ & $0.57\pm 0.04$ & $0.44\pm 0.04$ & $0.66\pm 0.04$ & $0.44\pm 0.04$ \\
$\beta R_{\rm eff}$ & $4.10 \pm 0.31$ & $5.07\pm 0.51$ & $4.81\pm 0.45$ & $5.66\pm 0.51$ & $5.71\pm 0.52$ & $5.34\pm 0.49$ & $5.47\pm 0.50$ \\
\hline
\end{tabular}
\caption{Fitted values for the parametrization of the spectral shape~\eqref{eq:fit} and fitted value of $\beta R_{\rm eff}$ in Eq.~\eqref{eq:efficiency}. The corresponding spectra are shown in Fig~\ref{fig:kappapfitplot}.}
\label{table:fit}
\end{table*}

For certain simple forms of the $f_l$ function the time integral in Eq.~\eqref{eq:Cpc} can be done analytically, which makes the simulation significantly faster. In particular, it can be done analytically if $f_l$ is a broken power-law with integer powers. We consider the form~\eqref{eq:kappaapr} for the efficiency factor $\kappa$ with $c=1$. Strictly speaking our results then hold for the case that $\Delta P_{1\to N} \propto \gamma$. However, since the difference in $\kappa(R)$ for $c=1$ and $c=2$ is very small, our results give a good approximation also of the case $\Delta P_{1\to N} \propto \gamma^2$. The pressure $\Delta P_{1\to N}$ mainly just determines the asymptotic radius $R_{\rm eq}$ through Eqs.~\eqref{eq:Req} and \eqref{eq:gammaeq}. In our simulations $R_{\rm eq}$ is an input parameter, and we perform the numerical simulations for several values of $R_{\rm eq}$. We also assume that $\mathcal{K}\approx 1$, which typically holds for strongly supercooled transitions, so that
\be \label{eq:efficiency}
\kappa(R) \approx \frac{1}{1+R/R_{\rm eq}} \,.
\ee

\section{Results}

\begin{figure*}
\centering
\includegraphics[width=0.92\textwidth]{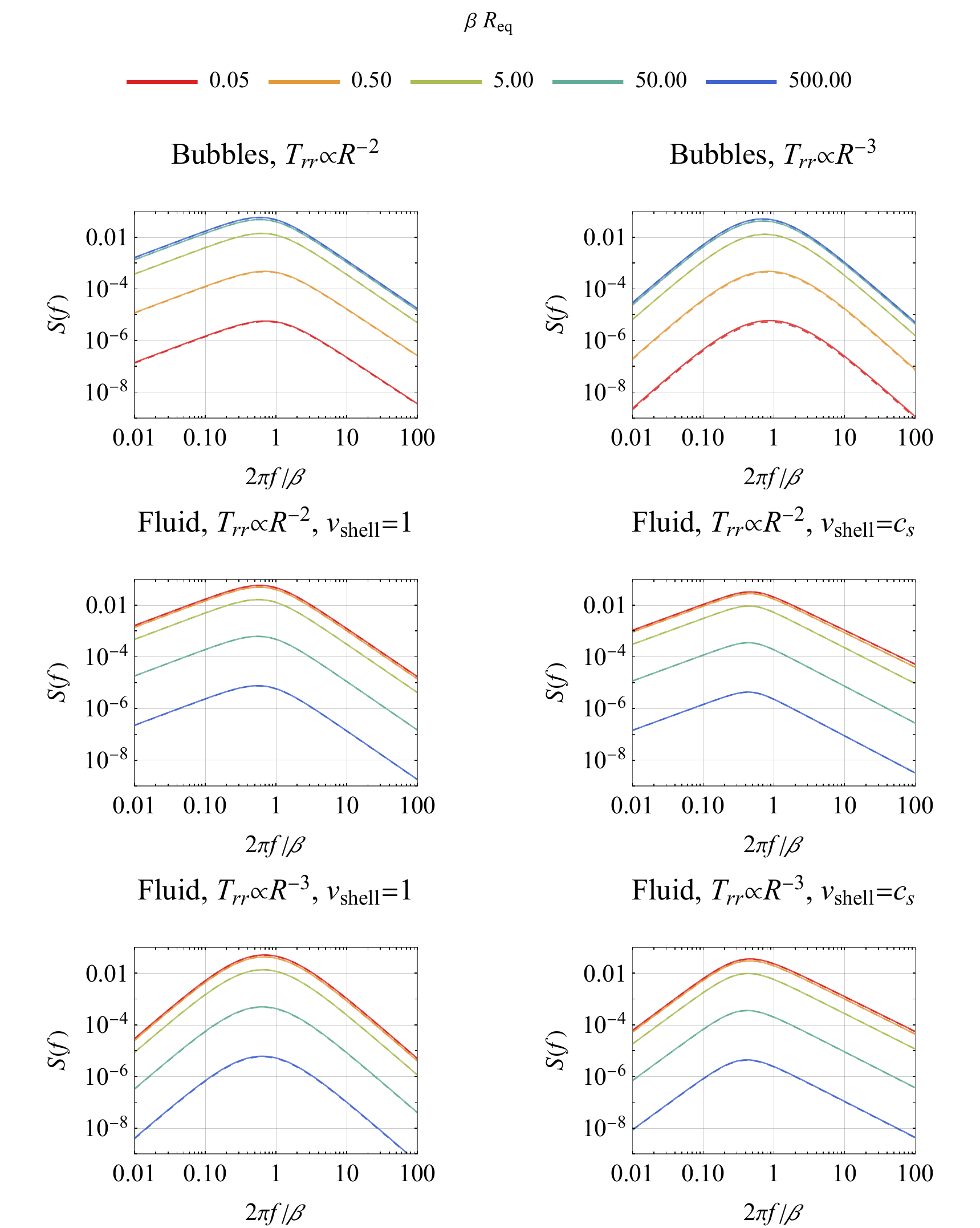}
\caption{Fitted GW spectral shape sourced by bubble walls and fluid motion assuming different scalings of the source after the collisions, $T_{rr} \propto R^{-\xi}$, and different velocities of the fluid shell after the collision. The solid curves show the results obtained by directly including the factor $\kappa(R)$ to the simulation and the dashed curves the result obtained by scaling the result obtained without it factor by $\kappa(R_{\rm eff})$.}
\label{fig:gws}
\end{figure*}

We perform 90 simulations with simulation volume $(16/\beta)^3$, each including at least 70 bubbles, for a range of $R_{\rm eq}$ values  including both signals due to bubble walls and the surrounding fluid in each of the cases described in the previous section. From the simulations we compute the spectral shape function~\eqref{eq:S}. In each case we fit the data combined from the 90 simulations with a broken power-law spectrum of the form
\be \label{eq:fit}
S_{\rm fit}(f) = \frac{A\,(a+b)^c}{\left[b \!\left(\frac{f}{f_p}\right)^{\!\text{-}\frac{a}{c}} \!+ a \!\left(\frac{f}{f_p}\right)^{\!\frac{b}{c}}\right]^c} \,,
\ee
where $a,b>0$ determine the low and high frequency power-law tails of the spectrum, $c>0$ the width of the transition between these power-laws, while $f_p$ and $A$ the peak frequency and amplitude of the spectrum respectively. The resulting GW spectra are shown in Fig.~\ref{fig:gws} with the solid curves. The color coding indicates different values of $R_{\rm eq}$.

\begin{figure}
\centering
\includegraphics[width=0.9\columnwidth]{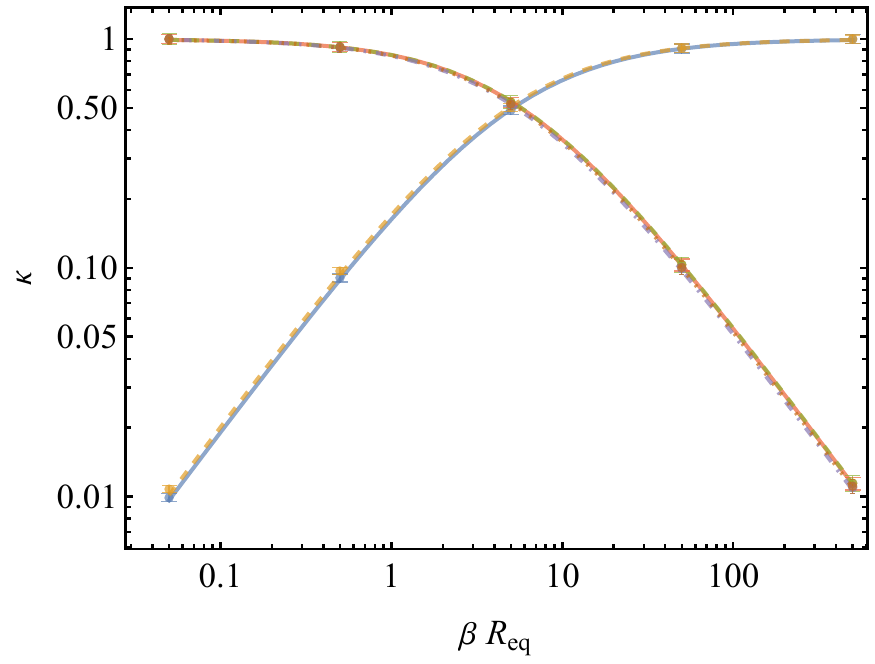}
\caption{The blue and red points with error bars show the amplitude of the GW spectrum from bubble collisions and from fluid motion, respectively, relative to the $R_{\rm eq}$ value that gives the largest amplitude. The blue and red curves show $\kappa_{\rm bubbles}$ and $\kappa_{\rm fluid}$, respectively. 
The parameters read $\xi=3$ and $v_{\rm fluid} = 1$ for both solid lines and the points of corresponding colour. Dashed and dotted lines and their corresponding points show the remaining cases ($\xi=2$ and $v_{\rm fluid}=c_s$) which as we see largely overlap with the previous two.}
\label{fig:kappapfitplot}
\end{figure}

For the solid curves in Fig.~\ref{fig:gws} the efficiency factor is directly included in the simulation as in Eqs.~\eqref{eq:Edecay} and \eqref{eq:Edecay2}. A commonly used approximation for the effect of the efficiency factor on the GW spectrum is to multiply the spectra obtained for bubble collisions and fluid motion without any efficiency factor by $\kappa(R_{\rm eff})^2$ and by $(1-\kappa(R_{\rm eff}))^2$, respectively. To check this, we have computed the amplitude of the GW spectrum in each case relative to the $R_{\rm eq}$ case that gives the largest amplitude and fitted $R_{\rm eff}$. The data points and resulting fits for all cases are shown in Fig.~\ref{fig:kappapfitplot} and the corresponding fitted values of $R_{\rm eff}$ in the last line of Table~\ref{table:fit}. We find that the effect of the efficiency factor is almost independent of the behaviour of the GW source after the collisions. In all cases our results give $R_{\rm eff} \simeq 5/\beta$, showing that the often used approximation with $R_{\rm eff} = (8\pi)^{1/3}/\beta \approx 2.9/\beta$ slightly underestimates $R_{\rm eff}$. Moreover, the results of applying the fitted efficiency factor are shown in Fig.~\ref{fig:gws} by the dashed curves. For these curves we have used the mean values given in Table~\ref{table:fit} that are obtained by averaging over the fits with different equilibrium radius, except for the amplitude for which we use the strongest signal for each source. We see that the dashed curves agree very well with the fully numerical results shown by the solid curves. This shows that the efficiency factor does not change the shape of the GW spectrum but gives only an overall suppression factor.

To summarize, we have shown that for very strong transitions, $\alpha\gg\alpha_\infty$, the GW spectrum from bubble collisions and from fluid motion, accounting for the distribution of energy between these sources, is given by
\be \label{eq:OmegaFit}
\!\!\Omega_{{\rm GW}}(f) \!=\! \left[\frac{H}{\beta}\right]^2 \!\left[\frac{\kappa(R_{\rm eff}) \, \alpha}{1+\alpha}\right]^2 \!\! \frac{A\,(a+b)^c}{\left[b \!\left(\frac{f}{f_p}\right)^{\!\text{-}\frac{a}{c}} \!+ a \!\left(\frac{f}{f_p}\right)^{\!\frac{b}{c}}\right]^c} \,,
\ee
where the efficiency factor is given by Eq.~\eqref{eq:efficiency}. The fitted values of the parameters $A, a, b, c, f_p$ and $R_{\rm eff}$ are given in Table~\ref{table:fit}. For weaker transitions, $\alpha\lesssim \alpha_\infty$, also the prefactor $\mathcal{K}$, given in Eq.~\eqref{eq:K}, needs to be accounted for, as well as the suppression arising from heating of the fluid around the bubble wall~\cite{Espinosa:2010hh}. In the limit of large wall velocity appropriate for strong transitions this reduction takes a simple form~\cite{Ellis:2019oqb}
\be
\kappa_{\rm fluid}=\frac{\alpha_{\rm eff}}{\alpha}\frac{\alpha_{\rm eff}}{0.73+0.083\sqrt{\alpha_{\rm eff}}+\alpha_{\rm eff}} \,,
\ee
where $\alpha_{\rm eff} = [1-\kappa(R_{\rm eff})]\alpha$.

The GW spectrum today can be obtained from~\eqref{eq:OmegaFit} by accounting for the scaling of the amplitude and frequency with the scale factor~\cite{Lewicki:2020jiv}:\footnote{Here for simplicity while red-shifting we assumed radiation dominated expansion from the transition time up to the matter-radiation equality. For a review of alternative scenarios and their impact on the spectra see Ref.~\cite{Allahverdi:2020bys}.}
\be
\begin{aligned}
&\Omega_{{\rm GW},0} = \frac{1.67\!\times\!10^{-5}}{h^2}  \!\left[\frac{100}{g_*}\right]^{\!\frac13}\! \Omega_{{\rm GW}}(f) \,, \\
& f_{p,0} = h_* \left[\frac{f_p}{\beta} \right] \left[\frac{\beta}{H}\right] \,,
\end{aligned}
\ee
where $h$ denotes the dimensionless Hubble constant, $h = 0.674$~\cite{Planck:2018vyg}, and $h_*$ the inverse Hubble time at the transition redshifted to its value today
\be
h_* = 1.65\times 10^{-5}\,{\rm Hz}\, \left[\frac{T_*}{100\,{\rm GeV}}\right] \left[\frac{g_*}{100}\right]^{\frac16} \,.
\ee 
Here $T_*$ denotes the temperature after the transition (including reheating) and $g_*$ the effective number of relativistic degrees of freedom at that temperature. At scales larger than the horizon scale at the time of the transition the source is not coherent and, consequently, in standard radiation domination the slope of the spectrum changes to $\Omega_{\rm GW}\propto f^3$ for $f < h_*$~\cite{Caprini:2009fx,Cai:2019cdl}.\footnote{The low frequency slope is also changed by possible modifications of the expansion rate~\cite{Barenboim:2016mjm,Hook:2020phx,Gouttenoire:2021jhk} although the only scenario in which the signal is not diminished is when the modification in question is itself caused by the transition for instance through slow decay of the scalar field leading to a period of matter domination~\cite{Ellis:2020nnr}.}

\section{Conclusions}

In this paper we have revisited the energy budget of strong first-order phase transitions to verify its impact on the produced gravitational wave spectra. We have gone beyond the current state-of-art by including the efficiency as a function of radius of the bubble accounting for the collision time of each point on the bubble surface. We have utilised numerical simulations randomly nucleating bubbles in a three dimensional box with periodic boundaries and used these to compute the GW spectra. This has allowed us to confirm that a simplified treatment of simply scaling entire spectra with an efficiency factor computed at some characteristic radius is accurate as the spectral shapes do not change due to the efficiency factor significantly. We did, however, find that in order to accurately describe the results the characteristic radius used in the simplified calculation should be around $R_{\rm eff}\approx 5/\beta$ rather than the usually employed average bubble separation $R_*= (8\pi)^\frac13 /\beta\approx 2.9 /\beta$.  

In each simulation we have also took into account the scaling of the GW sources after the collision in order to provide new fits for the resulting spectra from strongly supercooled transitions. The results are shown in Table~\ref{table:fit} and, starting from strongest transitions, include bubble collision spectra for both $T_{rr}\propto R^{-3}$, appropriate for gauge symmetry breaking, and $T_{rr}\propto R^{-2}$, appropriate for global symmetry breaking. Going towards slightly weaker transitions, we have provided the spectrum generated by fluid motion with the scaling $T_{rr}\propto R^{-3}$ and assuming the fluid remains in the form of relativistic shocks $v_{\rm fluid}=1$ after the transition. For transitions which are not extremely strong, we have show results closer to the sound wave picture in which the velocity of the fluid quickly relaxes to the speed of sound $v_{\rm fluid}=c_s$, again assuming the scaling $T_{rr}\propto R^{-3}$. Finally, for illustration, we also provide fluid spectra assuming the scaling $T_{rr}\propto R^{-2}$.  

Taking into account that for very relativistic walls the fluid profiles are extremely peaked, we have thus show that the final GW spectrum will be indistinguishable from an even stronger transition where bubble collisions would be the main source. Only for weaker transitions where the hydrodynamical effects change the propagation speed of the fluid shells, the spectrum diverges from the spectrum arising from bubble collisions.

\begin{acknowledgments}
This work was supported by the Spanish MINECO grants FPA2017-88915-P and SEV-2016-0588, the Spanish MICINN grants IJC2019-041533-I and PID2020-115845GB-I00/AEI/10.13039/501100011033, the grant 2017-SGR-1069 from the Generalitat de Catalunya, the Polish National Science Center grant 2018/31/D/ST2/02048, and the Polish National Agency for Academic Exchange within Polish Returns Programme under agreement PPN/PPO/2020/1/00013/U/00001. IFAE is partially funded by the CERCA program of the Generalitat de Catalunya.
\end{acknowledgments}

\bibliography{gw}

\end{document}